# Quantitative noncontact measurement of thermal Hall angle and transverse thermal conductivity by lock-in thermography

Takumi Imamura[1,2,3], Takamasa Hirai[1], Koichi Oyanagi[1,3], Ryo Iguchi[1], Kenta Takamori[3], Satoru Kobayashi[3], and Ken-ichi Uchida[1,2,4,*]

[1] National Institute for Materials Science, Tsukuba 305-0047, Japan

[2] Graduate School of Science and Technology, University of Tsukuba, Tsukuba 305-8573, Japan

[3] Faculty of Science and Engineering, Iwate University, Morioka 020-8551, Japan

[4] Department of Advanced Materials Science, Graduate School of Frontier Sciences,
The University of Tokyo, Kashiwa 277-8561, Japan

[*] UCHIDA.Kenichi@nims.go.jp

**Abstract**

We propose and demonstrate a quantitative noncontact measurement method for the thermal Hall effect (THE) based on magnetic-field-modulated lock-in thermography. This method enables visualization of THE-induced temperature change and quantitative estimation of the thermal Hall angle $\theta_{\text{THE}}$ by applying periodic magnetic fields to a sample and obtaining the first harmonic response of thermal images. By combining this method with LIT-based measurement techniques for the longitudinal thermal conductivity $\kappa_{xx}$, we also quantify the transverse thermal conductivity $\kappa_{xy}$. We validate our measurement methods by estimating $\theta_{\text{THE}}$, $\kappa_{xx}$, and $\kappa_{xy}$ in a ferromagnetic Heusler alloy $Co_2MnGa$ slab showing large THE.

## I. INTRODUCTION

The thermal Hall effect (THE), known as the Righi–Leduc effect, is the thermal counterpart of the Hall effect [1-15]. THE generates the transverse heat current density $\mathbf{j}_q^{\text{THE}}$ in the direction of the cross product of the input heat current density $\mathbf{j}_q^{\text{in}}$ and magnetic field $\mathbf{H}$ (with the magnitude $H$) or magnetization $\mathbf{M}$ (with the magnitude $M$) as

$$\mathbf{j}_q^{\text{THE}} = \frac{\kappa_{xy}}{\kappa_{xx}} \mathbf{j}_q^{\text{in}} \times \frac{\mathbf{H}}{|H|} \left(\text{or } \frac{\mathbf{M}}{|M|}\right) = \theta_{\text{THE}} \, \mathbf{j}_q^{\text{in}} \times \frac{\mathbf{H}}{|H|} \left(\text{or } \frac{\mathbf{M}}{|M|}\right), \quad (1)$$

where $\kappa_{xy}$ is the transverse thermal conductivity, $\kappa_{xx}$ the longitudinal thermal conductivity, and $\theta_{\text{THE}} = \kappa_{xy}/\kappa_{xx}$ the thermal Hall angle [9,13,15]. Since various heat carriers in solids, including electrons, phonons, and magnons, can induce THE, its measurement provides a crucial piece of information to investigate transport properties, e.g., the effects of spin-orbit interaction and Berry curvature on the heat carriers [1–13]. Finding materials showing large THE is also important for the thermal management because THE enables active heat flow control using a magnetic field and/or magnetic materials [14]. THE has been measured by attaching temperature sensors directly to samples. This conventional method requires trained experimental skills to minimize heat leakage from the samples to the temperature sensors, a situation which makes it difficult to ensure the reliability and reproducibility of the experiment. Therefore, a simple and noncontact measurement method for THE needs to be established.

A method to realize noncontact measurements of THE is the lock-in thermography (LIT) [15]. LIT is an active thermal imaging technique that detects a thermal response excited by a periodic external perturbation applied to a sample with high-temperature resolution [14–36]. The LIT-based THE measurement was proposed by Tomioka *et al.* [15], where they applied periodic laser heating with an elliptical spot shape to a Bi slab under a magnetic field and obtained thermal images synchronized with the frequency of the laser heating. In this method, the LIT results include both longitudinal and transverse thermal conduction contributions. Thus, when the contribution of longitudinal thermal conduction is dominant, it is difficult to visualize the THE-induced thermal signals and to quantitatively estimate the thermal Hall angle.

In this study, we propose and demonstrate a noncontact quantitative measurement method for THE. Based on magnetic-field-modulated LIT, we quantified $\theta_{\text{THE}}$ in a ferromagnetic Heusler alloy Co$_2$MnGa (CMG), which is known to exhibit large THE [7]. We also determined $\kappa_{xy}$ of CMG by combining the obtained $\theta_{\text{THE}}$ value with a LIT-based measurement of $\kappa_{xx}$; all the coefficients related to THE can be determined by LIT in a single setup. The $\theta_{\text{THE}}$ and $\kappa_{xy}$ values obtained by our LIT-based measurements are consistent with those reported in the literature, confirming the validity of our method.

This paper is organized as follows. In Sec. II, we explain the measurement principle and experimental procedure for the thermal imaging of THE based on magnetic-field-modulated LIT. In Sec. III, we demonstrate the thermal imaging of THE and quantitative estimation of $\theta_{\text{THE}}$ and $\kappa_{xy}$ using the CMG slab. Section IV is devoted to the conclusion of the present study.

## II. MEASUREMENT PRINCIPLE

The measurement principles and processes of the quantitative estimation of THE using the LIT method are illustrated in Fig. 1. For the measurement of THE, we adopted the magnetic-field-modulated LIT method, where a periodic magnetic field is applied to a sample and thermal images synchronized with the frequency of the field are extracted. We quantified the THE-induced temperature modulation and the value of $\theta_{\text{THE}}$ by performing magnetic-field-modulated LIT with applying a steady temperature gradient along the *x* direction (left flow chart in Fig. 1). To determine $\kappa_{xy}$, $\kappa_{xx}$ must be known. The estimation of $\kappa_{xx}$ is realized by measuring the thermal diffusivity $D$ and volumetric specific heat $\rho c_p$ through the relation $\kappa_{xx} = D\rho c_p$, where $\rho$ and $c_p$ represent the density and specific heat, respectively. The LIT method also enables the determination of $D$ and $\rho c_p$ through the measurements of the thermal diffusion and Joule heating with applying a periodic heat current (center flow chart in Fig. 1) and a periodic charge current (right flow chart in Fig. 1), respectively [30,32]. By combining $\theta_{\text{THE}}$ and $\kappa_{xx}$ obtained by the above experiments, we quantified $\kappa_{xy}$. These procedures allow us to estimate all phenomenological parameters related to THE without attaching temperature sensors to the sample.

Figure 2 shows the details of the magnetic-field-modulated LIT measurement of THE. In this method, we measured thermal images of a sample surface with applying periodic magnetic fields with the first harmonic amplitude $\mu_0 H_{1f}$, frequency $f$, and zero offset [Fig. 2(a)]. We extracted the first harmonic response of the detected images, which are transformed into the lock-in amplitude $A$ and phase $\varphi$ images through Fourier analysis. The $\varphi$ image contains information about the sign of the temperature change and time delay associated with the thermal diffusion. We note that the waveform of the applied periodic magnetic field is not perfectly sinusoidal. This is because the square wave reference signal from the LIT system is passed through a low-pass filter and then amplified to generate a periodic current in the electromagnet by using a bipolar current source. Therefore, the waveform of the periodic magnetic field was measured directly by a Hall sensor. From the measured periodic field, the first harmonic amplitude $\mu_0 H_{1f}$ and phase are extracted through Fourier analysis. The phase values are subtracted from the raw $\varphi$ images to synchronize the thermal response with the first harmonic magnetic field oscillations.

Figure 2(b) shows a schematic illustration of the experimental setup for the measurements of THE [29]. This setup is designed to demonstrate the symmetry of THE in a single LIT image. We attached a chip heater to the center of a bar-shaped sample using instant adhesive (Bond Aron Alpha, Konishi Co., Ltd.). Both ends of the sample were thermally connected to heat baths consisting of a Cu plate, where both Cu plates were attached to the sample by soldering. As shown in the following section, the heater power $P$ is not necessary to quantify THE; thus, our sample setup does not require careful attachment of the chip heater to minimize its heat loss. This ensures the simplicity and high reproducibility of our experiments. By applying a charge current to the heater with $P$, the static temperature gradient $\nabla T$ was generated in the sample, where the direction of $\nabla T$ in the region R1 is opposite to that in the region R2; the direction of $\mathbf{j}_q^{\text{in}}$ is reversed across the center of the sample along the $x$ direction [Fig. 2(b)]. During the LIT measurements, we applied a uniform and periodic magnetic field to this sample. Thus, based on the symmetry of THE [see Eq. (1)], the sign of the temperature change induced by THE is expected to be reversed across the heater. For precise confirmation of the symmetry of THE, we performed the LIT measurements in both the out-of-plane and in-plane magnetic field configurations [Figs. 2(c) and 2(d), respectively]. In the out-of-plane (in-plane) magnetic field configurations, the transverse heat current due to THE is generated along the $z$ ($y$) direction. To enhance the infrared emissivity, the surface of the sample was coated with insulating black ink. The calibration to convert infrared radiation intensity into temperature was performed using a temperature sensor coated with black ink under the same conditions as the LIT measurements (see Ref. [16] for details). In the LIT measurements, THE was measured at the lock-in frequencies ranging from 1 to 5 Hz, where possible signal reduction due to the thermal diffusion in the black-ink layer is negligibly small [37]. The black ink does not promote the corrosion and oxidation on the surface of the samples. We carried out all the measurements at room temperature and atmospheric pressure. We defined the coordinate so that the definition of $\kappa_{xy}$ in the in-plane magnetic field configuration is consistent with that of the literature [9,13]. The absolute value of $\theta_{THE}$ is estimated by taking the average of the analyzed results in R1 and R2 in the in-plane magnetic field configuration. Note that, although the periodic magnetic field was used as the input signal of LIT for the direct measurement of the magnetocaloric effect [18,35], our method can exclude the contamination of this phenomenon because its contribution appears in the second-harmonic component in our measurement condition [36]. In contrast, in this study, we detect only the first-harmonic component including the THE-induced signal.

## III. EXPERIMENTAL DEMONSTRATION

### A. Sample preparation

We prepared polycrystalline CMG slabs by the spark plasma sintering method from a CMG ingot. The detailed recipe for the CMG ingot is described in Ref. [38]. The CMG ingot was crushed using a mortar and planetary ball mill, followed by sieving through a 100 μm mesh. Next, a cylindrical CMG slab with a diameter of 20 mm was produced by sintering the prepared CMG powder twice; the first (second) sintering was done at 831 °C (895 °C) and 30 MPa for 30 min. The sintered CMG slab was cut into rectangular samples with a $15.0 \times 1.0 \times 0.7$ mm dimension. For comparison, we also used ferromagnetic Ni, $Ni_{70}Pd_{30}$, and $Ni_{25}Fe_{75}$ slabs prepared by a melting method with rapid cooling, available from Kojundo Chemical Laboratory Co., Ltd., Japan, which were cut into a rectangular shape with a dimension of $15.0 \times 1.0 \times 0.5$ mm. These samples are identical to those used in Ref. [39]. These rectangular samples were used for the LIT-based THE measurement and magnetization measurement using a vibrating sample magnetometry.

### B. Quantitative determination of thermal Hall angle using LIT

We carried out LIT measurements under low magnetic fields, e.g., $\mu_0 H_{1f} < 50$ mT, where the magnetization and THE-induced signals depend linearly on the magnetic fields [Figs. 3(a) and 3(b)].

In this regime, we can quantify THE since our method detects linear response contributions with respect to periodic magnetic fields (note that our method can be extended to a non-linear response regime in principle by using the multi-harmonic detection technique [32,36]). Figures 3(c) and 3(d) [3(e) and 3(f)] respectively show the $A$ and $\varphi$ images of the CMG slab at $f$ = 2.0 Hz, $P$ = 76 mW, and $\mu_0 H_{1f}$ = 43 mT (47 mT) in the out-of-plane magnetic field (in-plane magnetic field) configuration, where $\mu_0$ is the vacuum permeability. In the out-of-plane magnetic field configuration, temperature changes appear near the sample edges [Fig. 3(c)], while the phase difference between the left and right halves is almost 180° [Fig. 3(d)]. This indicates that the sign of the temperature change on the left half is opposite to that on the right half in the CMG slab. The amplitude of the temperature changes varies almost linearly in the $z$ direction in the right and left halves of the CMG slab, which is consistent with the temperature distribution induced by the transverse heat current in a steady state. The sign of the temperature change is reversed across the heater position, indicating that the observed phenomenon depends on the direction of $\mathbf{j}_q^{in}$ and is consistent with the symmetry of THE. This interpretation is further confirmed by the LIT images in the in-plane magnetic field configuration. As shown in Fig. 3(e), almost uniform temperature changes appear on the CMG surface except for the area near the heater, where a temperature gradient is absent. The phase difference between the R1 and R2 regions is almost 180° [Fig. 3(f)], indicating that the sign of the temperature change is reversed across the heater position. We confirmed that the magnitude of the temperature changes is proportional to $\nabla T$ applied to the CMG slab [Fig. 3(g)], where $\nabla T$ is estimated by fitting the steady-state temperature profile in R1 and R2 with a linear function [Fig. 2(b)]. These results are in good agreement with the expected behavior of THE. Importantly, we observed no signal in the absence of a temperature gradient [Fig. 3(g)], confirming that our method enables the pure detection of THE.

For quantitative discussion of THE, we need to estimate the magnitude of the THE-induced temperature change in the steady state. In general, LIT images measured at low $f$ show temperature distributions in nearly steady states, while those at high $f$ show temperature distributions in transient states [25,28]. Thus, we measured the $f$ dependence of the THE-induced temperature change [Fig. 3(h)]. Here, the THE-induced signals in Fig. 3(h) are measured at $\mu_0 H_{1f} \sim 50$ mT, but the values of $\mu_0 H_{1f}$ at various frequencies are slightly different. Thus, the magnitude of the THE-induced signals in Fig. 3(h) are normalized by each value of $\mu_0 H_{1f}$ so that the difference of $\mu_0 H_{1f}$ at various frequencies does not affect quantifying THE (note that the THE signal is proportional to $\mu_0 H_{1f}$ in such a low field range). Figure 3(h) shows that the normalized amplitude of the THE-induced temperature changes increases with decreasing $f$. This behavior can be explained by the one-dimensional heat diffusion. We estimated the normalized THE-induced signals $\alpha$ [$\equiv A/(\mu_0 H_{1f} \cdot \nabla T)$] in the steady state by fitting the $f$ dependence of $\alpha$ with the solution of the one-dimensional heat diffusion equation [25, 28].

The steady-state magnitude of the THE-induced temperature change enables the estimation of the thermal Hall angle of the CMG slab. Here, we assume that obtained $\alpha$ is dominated by the magnetization-dependent anomalous THE, while the contribution of the magnetic-field-dependent ordinary THE is negligibly small. This is validated based on the previous experimental reports on THE for CMG [7], where the magnetic field dependence of $\kappa_{xy}$ shows that the ordinary component in the single-crystalline CMG slab is estimated to be less than 1.8% of the total THE-induced signals at around 50 mT, i.e., the magnetic field range of our LIT measurement. This assumption allows us to estimate the THE-induced temperature change at the saturation field $\Delta T^{\text{THE}}$ using the magnetization process: $\Delta T^{\text{THE}}/\nabla T = 2\alpha^{\text{steady}}(\partial H/\partial M)M_s$, where $M_s$ is the saturation magnetization, $\alpha^{\text{steady}}$ is the steady-state value of $\alpha$, and $(\partial H/\partial M)M_s$ means the saturation field. This estimation is based on the fact that the magnetic field dependence of the THE-induced signal is proportional to the magnetization process. From $\Delta T^{\text{THE}}$, we can estimate $\theta_{\text{THE}}$ as

$$\theta_{\text{THE}} = \frac{\Delta T^{\text{THE}}}{t\nabla T}, \qquad (2)$$

where $t$ is the thickness of the CMG slab (= 0.7 mm). By applying the above analysis procedures to R1 and R2 and taking the average of their values, we estimated $\theta_{\text{THE}}$ of the CMG slab to be +2.94±0.09%, which is comparable to that in the previous work, i.e., 2.65% in the single crystalline CMG slab [7].

## C. Measurement of longitudinal and transverse thermal conductivities using LIT

Now we are in a position to estimate $\kappa_{xy}$. To do this, we performed heat-current-modulated LIT (LIT measurement 1) and charge-current-modulated LIT (LIT measurements 2 and 3) to estimate $D$ and $\rho c_p$, respectively, enabling the estimation of $\kappa_{xx}$ by LIT (Figs. 1 and 4). By combining the measured $\theta_{\text{THE}}$ and $\kappa_{xx}$ values, we can determine $\kappa_{xy}$ only by using LIT measurements. In the LIT measurement 1, the LIT system inputs a periodic heat current to the CMG slab by applying square-wave-modulated $P$ to the heater and detects thermal images oscillating with the same frequency as the heat current. These LIT images show the thermal diffusion along the $x$ direction induced by the periodic heating, which can be reproduced by the calculation based on the one-dimensional heat diffusion equation [30]. Thus, we can estimate $D$ by fitting the $x$-directional line profile of $\varphi$ with a linear function. In the LIT measurement 2 (LIT measurement 3), when we apply the square-wave-modulated charge current $\mathbf{J}_c$ with the amplitude $J_c$ and zero DC offset (the amplitude $J_c/2$ and DC offset $J_c/2$) to the sample, the LIT images show the temperature change due to the Peltier effect generated around electrodes placed at the ends of the slab (both the Joule heating in the bulk of the CMG slab and the Peltier effect at the ends of the slab). Because the thermal imaging of the Joule heating gives information on $\rho c_p$, we extracted pure Joule-heating-induced LIT signals with the amplitude $A_J$ and phase $\varphi_J$ by subtracting half of the LIT signals obtained in the LIT measurement 2 from those in the LIT measurement 3 [30]. From $A_J$, we can determine $\rho c_p$ as

$$\rho c_p = \frac{j_c^2}{A_J f \pi^2 \sigma}, \qquad (3)$$

where $\sigma$ is the electrical conductivity and $j_c$ is the first harmonic amplitude of the square-wave-modulated charge current density in the LIT measurement 3 [30]. To determine $\rho c_p$, the value of $\sigma$ measured by the four-probe method is substituted into Eq. (3). To check the validity of our method, we also measured $\kappa_{xx}$ of the CMG slab using the conventional methods; $\kappa_{xx}$ was determined by combining the values of $D$ measured by the laser flash method, $c_p$ by the differential scanning calorimetry, and $\rho$ by the Archimedes method. Note that the measurements and analysis of the charge-current-modulated LIT could be more simplified by placing the electrodes further away from the viewing area of the camera using a longer sample, in which raw data obtained from the LIT measurements 3 include the pure Joule-heating contribution free from the Peltier-effect-induced temperature change. This allows us to omit the LIT measurement 2 because its role is just estimating the contamination of the Peltier effect in the LIT measurement 3, which appears around the position of the electrodes, i.e., the ends of the sample, minimizing the complexity and improving the usability of the estimation of $\rho c_p$. Nevertheless, to confirm the pure detection of the Joule-heating signals, we adopted the procedures depicted in Fig. 4. The design of the sample dimensions is also important to ensure the one-dimensional heat diffusion in R1 and R2, leading to accurate determination of $D$.

In accordance with the above LIT-based experiments and analysis, we obtained $\kappa_{xx}$ from $D$ and $\rho c_p$ shown in Fig. 5, followed by the estimation of $\kappa_{xy}$. Figures 5(a) and 5(b) show the $A$ and $\varphi$ images and corresponding $x$-directional line profiles averaged over the slab measured by heat-current-modulated LIT. The $A$ ($\varphi$) line profile exhibits a symmetric exponential decay (linear change) with a distance from the center of the sample, i.e., the heater position. The observed behaviors indicate that $A$ and $\varphi$ originate from the thermal diffusion by modulating $P$ [30,32]. To derive $D$, the $x$-directional line profiles of $\varphi$ in R1 and R2 are fitted by linear functions [Fig. 5(b)]. The obtained $D$ shows a good agreement with that measured by the conventional method [see the dashed line in Fig. 5(c)]. Figures 5(d) and 5(e) show the $A_J$ and $\varphi_J$ images and corresponding $x$-directional line profiles averaged over the slab measured by charge-current-modulated LIT. The $A_J$ and $\varphi_J$ images show almost a uniform distribution of temperature change in the slab, which is consistent with the expected behavior of Joule heating. From the averaged $A_J$ values over the dashed rectangle area in Fig. 5(d), we estimated the values of $\rho c_p$ in the CMG slab, which are consistent with those measured by the conventional methods [see the dashed line in Fig. 5(f)]. By taking the average of the $D$ and $\rho c_p$ values at various frequencies and multiplying them, we estimated $\kappa_{xx}$ to be 20.8±0.7 W/mK, which is also in good agreement with that obtained by the conventional methods: 20.6 W/mK. Finally, by using the $\theta_{THE}$ and $\kappa_{xx}$ values obtained by our methods, we quantified $\kappa_{xy}$ to be 0.60±0.03 W/mK (note that $\kappa_{xy}$ of single-crystalline CMG reported in the previous work is 0.57 W/mK [7]). This consistency demonstrates that our LIT-based methods are valid for the quantitative measurement of the THE-related transport parameters.

### D. Demonstration of THE measurements for various ferromagnetic metals

We performed THE measurements on various ferromagnetic metals, which are expected to exhibit smaller THE-induced signals, to demonstrate the versatility of our measurement technique. To achieve this, we employed the same LIT-based experiment and analysis methods for the Ni, $Ni_{70}Pd_{30}$, and $Ni_{25}Fe_{75}$ slabs. The analysis areas, R1 and R2, are defined in the $A$ image for these metals [Fig. 6(g)]. The THE measurements were performed under $\mu_0 H_{1f}$ < 50 mT, where THE-induced signals show a linear response to $\mu_0 H_{1f}$, to quantify THE by magnetic-field-modulated LIT [Fig. 6(a)-6(c)]. Figures 6(d)-6(f) [6(g)-(i)] respectively show the $A$ and $\varphi$ images for the Ni, $Ni_{70}Pd_{30}$, and $Ni_{25}Fe_{75}$ slabs at $f$ = 2.0 Hz and $\mu_0 H_{1f}$ = 43 mT (47 mT) in the out-of-plane magnetic field (in-plane magnetic field) configurations, where the heater power is set at $P$ = 202 mW for Ni, 96 mW for $Ni_{70}Pd_{30}$, and 138 mW for $Ni_{25}Fe_{75}$. The $A$ images in the out-of-plane magnetic field (in-plane magnetic field) configurations show that temperature changes appear only in the vicinity of the edges of the slabs (uniformly on the surface of the slabs except for the center area near the heater). The $\varphi$ images show the signs of the temperature changes are reversed across the heater position. These results indicate that the observed signals are due to THE. Notably, the $\varphi$ values for $Ni_{25}Fe_{75}$ differ by 180° from those for other metals, reflecting the difference in the signs of $\theta_{THE}$, which is consistent with the behaviors expected from the anomalous Hall angles of the metals [39]. Figure 6(j) shows that the magnitude of the temperature changes varies linearly with respect to applied $\nabla T$, verifying that pure THE-induced signals are detected without the contamination of the magnetocaloric effect.

To determine the $\theta_{THE}$ values of Ni, $Ni_{70}Pd_{30}$, and $Ni_{25}Fe_{75}$, the $f$ dependences of $\alpha$ are derived using the results in the in-plane magnetic field configuration. Figure 6(k) represents the $f$ dependence of $\alpha$ in R1. The $f$-independent $\alpha$ signals indicate that the THE-induced signals at 1-5 Hz in these ferromagnetic metals reach the steady states, while the $f$ dependence of $\alpha$ in the CMG slab shows the slight $f$-dependent variation as shown in Fig. 3(h). The behavior of the $f$ dependence of $\alpha$ depends on the sample dimension along the $\mathbf{j}_q^{THE}$ direction, i.e., the thickness of the samples in the in-plane magnetic field configuration, and the thermal properties of the materials, i.e., $\kappa_{xx}$ and $D$ [25, 28]. The $\kappa_{xx}$ and $D$ values of Ni, $Ni_{70}Pd_{30}$, and $Ni_{25}Fe_{75}$ are higher than those of CMG and their thickness (0.5 mm) is smaller than that of CMG (0.7 mm). Therefore, the THE-induced signals for Ni, $Ni_{70}Pd_{30}$, and $Ni_{25}Fe_{75}$ reach the steady states quickly in the above $f$ range. Despite the substantially different thermal properties, the LIT-based measurements enable quantification of $\theta_{THE}$, as shown in Fig. 6 (l). The $\theta_{THE}$ value for Ni is in good agreement with the literature value [4], confirming that our technique quantifies $\theta_{THE}$ even for the materials with modest THE. This demonstrates the feasibility of applying this technique to a broad range of materials.

## IV. CONCLUSION

We proposed and demonstrated the quantitative noncontact measurement of THE based on the LIT technique. By means of magnetic-field-modulated LIT, we visualized the spatial distribution of the temperature changes induced by THE and quantified $\theta_{THE}$ of the polycrystalline CMG slab. Moreover, we determined $\kappa_{xy}$ by performing the noncontact estimation of $\kappa_{xx}$ through the heat-current-modulated and charge-current-modulated LIT methods. These parameters obtained by our methods are quantitatively consistent with those obtained by the conventional methods. Furthermore, the versatility of our method was demonstrated by measuring THE for the ferromagnetic Ni, Ni$_{70}$Pd$_{30}$, and Ni$_{25}$Fe$_{75}$ slabs showing smaller THE-induced signals than CMG. The estimated $\theta_{THE}$ value of Ni is also consistent with the literature value [4]. Although our method enables us to quantify THE only in the linear response regime at present, there are a lot of intriguing materials showing non-linear THE response to magnetic field [4–13]. Thus, it is desirable for our method to be integrated into multi-harmonic detection techniques for further accelerating the fundamental physics of THE. It could be one of the future works. Because our method is applicable to various materials, including metals, semiconductors, and insulators, it will potentially stimulate further materials science studies of this phenomenon.

**Acknowledgments**

The authors thank M. Isomura and K. Suzuki for technical support. This work was partially supported by ERATO "Magnetic Thermal Management Materials" (JPMJER2201) from JST, Japan; Grant-in-Aid for Scientific Research (S) (22H04965), Grant-in-Aid for Research Activity Start-up (22K20495), and Grant-in-Aid for Early-Career Scientists (21K14519) from JSPS KAKENHI, Japan; NEC Corporation; and NIMS Joint Research Hub Program.

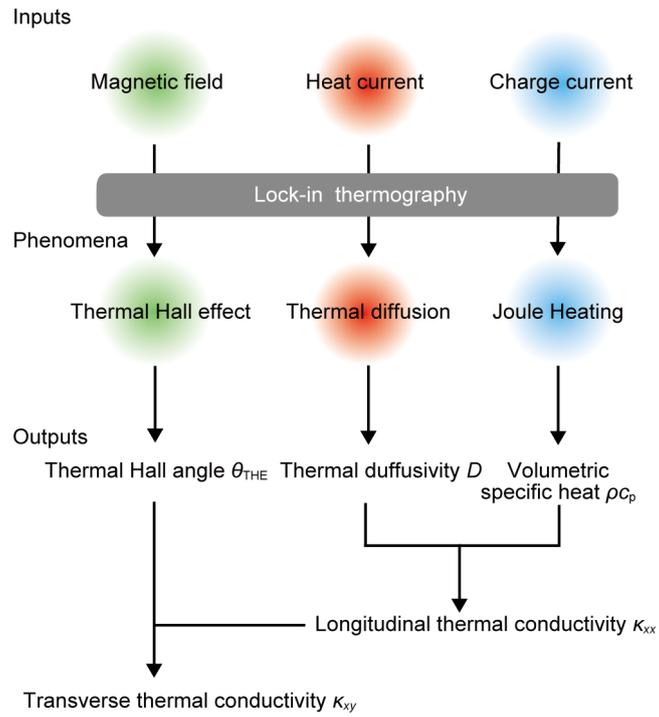

**FIG. 1.** Schematic diagram of the procedure for estimating the transverse thermal conductivity $\kappa_{xy}$ using the LIT technique. Through the magnetic-field-modulated LIT measurements, the thermal Hall angle $\theta_{\text{THE}}$ is estimated. $\kappa_{xy}$ is determined as $\theta_{\text{THE}}\kappa_{xx}$ with $\kappa_{xx}$ being the longitudinal thermal conductivity. $\kappa_{xx}$ estimated by multiplying the thermal diffusivity $D$ and volumetric specific heat $\rho c_p$, which are measured by heat-current-modulated and charge-current-modulated LITs, respectively.

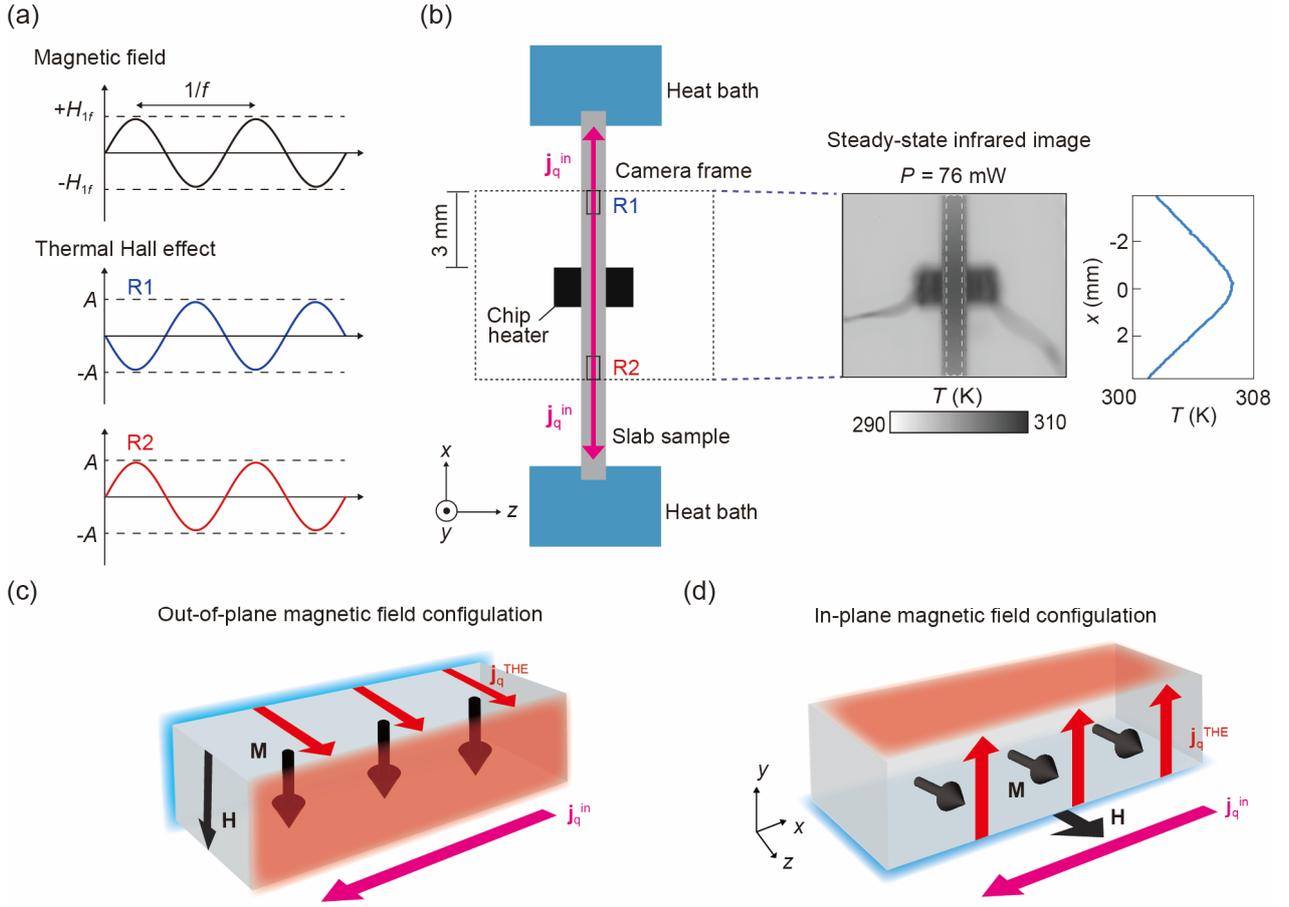

**FIG. 2.** (a) Input and output of magnetic-field-modulated LIT. A periodic magnetic field with the first harmonic amplitude $H_{1f}$, frequency $f$, and zero offset was applied, and the first harmonic response of thermal images was detected through LIT. (b) Schematic illustration of the experimental setup for the measurement of THE and steady-state temperature $T$ distribution with the heater power $P = 76$ mW. The input heat current density $\mathbf{j}_q^{in}$ flows in the opposite direction from the heater to the ends of the sample, and the resultant temperature gradient $\nabla T$ is reversed across the heater. The regions R1 and R2 show the analysis areas of the LIT images. (c), (d) Schematic illustrations of THE in the out-of-plane (c) and in-plane (d) magnetic field configulations. $\mathbf{j}_q^{THE}$, $\mathbf{H}$, and $\mathbf{M}$ denote the transverse heat current density induced by THE, magnetic field, and magnetization, respectively.

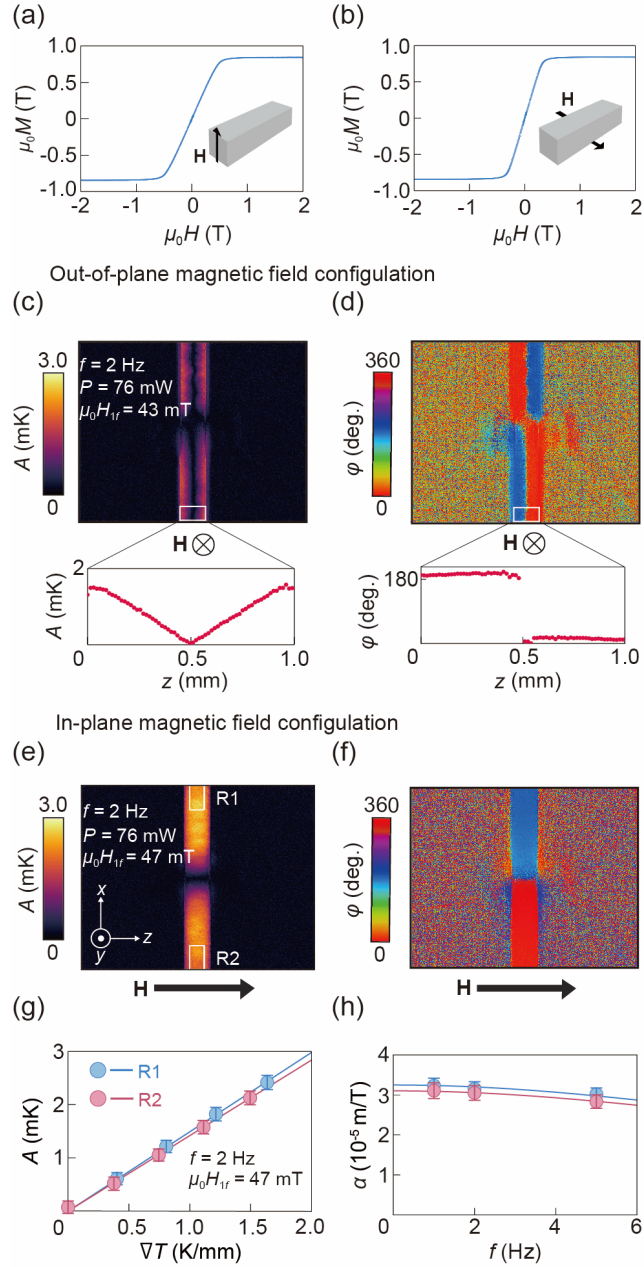

**FIG. 3.** (a), (b) Magnetization $M$ curves of the CMG slab in the out-of-plane (a) and in-plane (b) magnetic field configurations. In (a) [(b)], the magnetic field was applied along the 0.7 mm (1.0 mm) direction of the CMG slab. (c), (d) Lock-in amplitude $A$ and phase $\varphi$ images for the CMG slab at $f = 2.0$ Hz, $P = 76$ mW, and $\mu_0 H_{1f} = 43$ mT in the out-of-plane magnetic field configuration. The graphs below the LIT images are the line profiles of $A$ and $\varphi$ along the $z$ direction. (e), (f) $A$ and $\varphi$ images for the CMG slab at $f = 2.0$ Hz, $P = 76$ mW, and $\mu_0 H_{1f} = 47$ mT in the in-plane magnetic field configuration. (g) $\nabla T$ dependence of $A$ in the regions R1 and R2 in the in-plane magnetic field configuration at $f = 2.0$ Hz and $\mu_0 H_{1f} = 47$ mT. (h) $f$ dependence of the normalized temperature change induced by THE $\alpha$ [$\equiv A/(\mu_0 H_{1f} \cdot \nabla T)$]. The solid lines show the calculated $f$ dependence of $\alpha$ of CMG based on the one-dimensional heat diffusion.

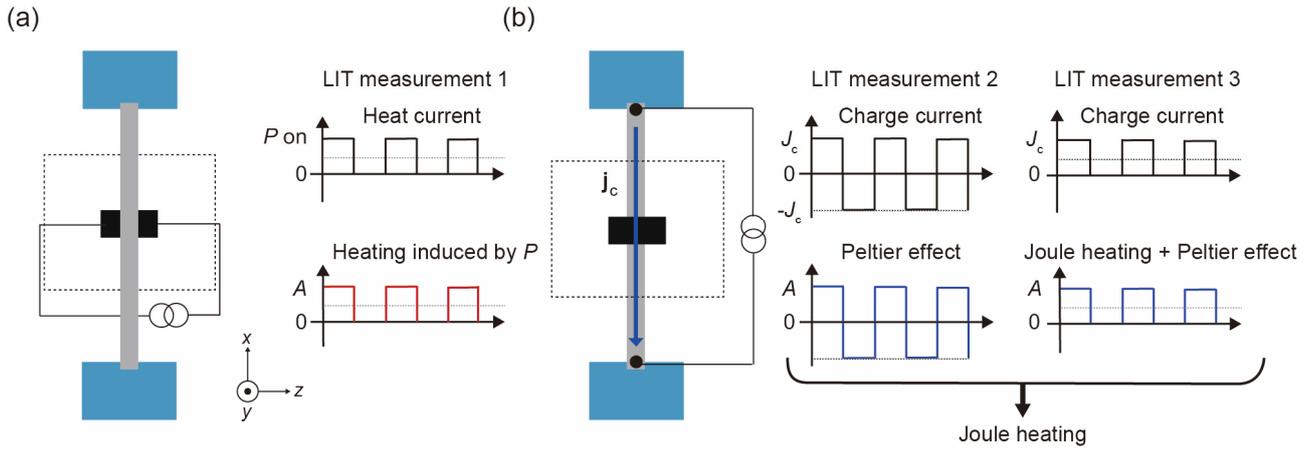

**FIG. 4.** Experimental configuration and procedure for estimation of $\kappa_{xx}$ using LIT. In the LIT measurement 1 (a), when we input a periodic charge current to the heater, i.e., a periodic heat current to the sample, we obtain the $A$ and $\varphi$ images reflecting the thermal diffusion from the heating at the center of the sample. In the LIT measurements 2 (3) (b), when we input a square-wave-modulated charge current with $f$, the amplitude $J_c$ ($J_c/2$), and zero offset (DC offset $J_c/2$) to the sample, we measure the Peltier effect at the end of the sample (the Joule heating and the Peltier effect at the end of the sample) in the LIT images. Using these LIT images, we extracted the $A_J$ and $\varphi_J$ images showing only the Joule heating contribution. We can determine $D$ and $\rho c_p$ from the spatial profile of $\varphi$ in the LIT measurement 1 and of $A_J$, respectively, resulting in the estimation of $\kappa_{xx}$.

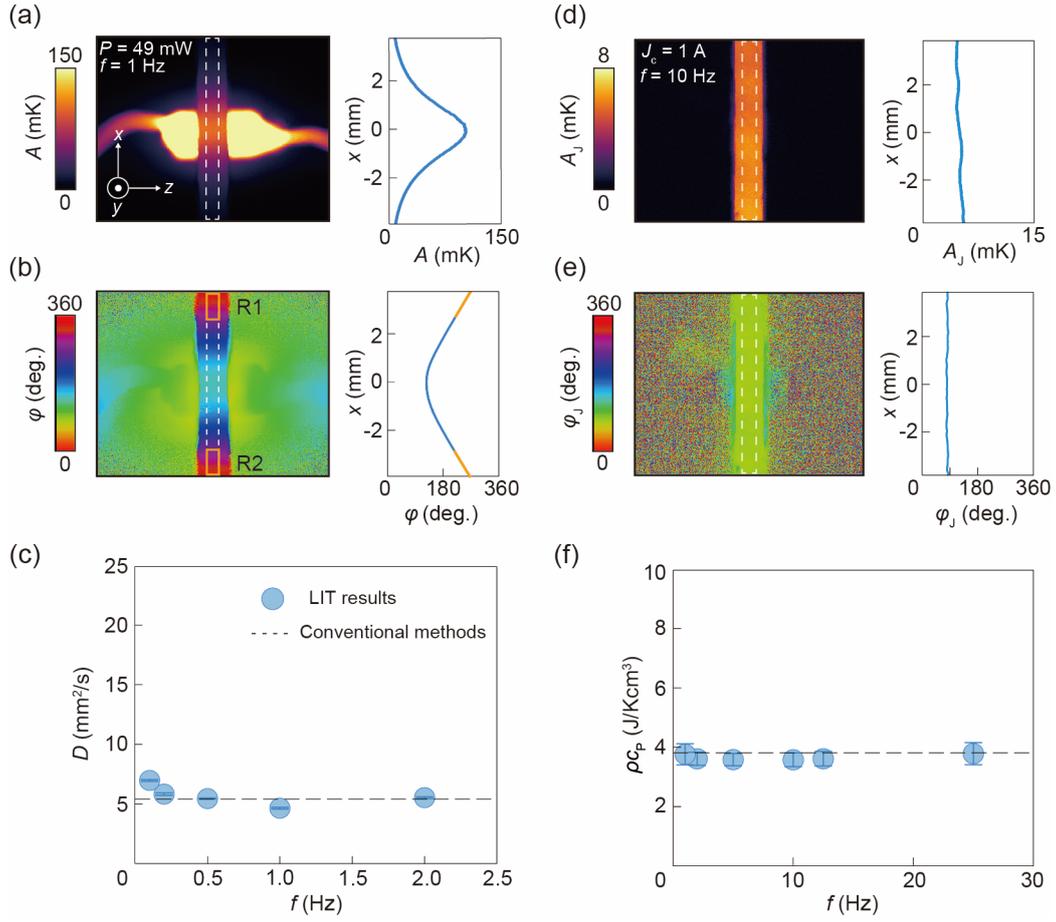

**FIG. 5** (a), (b) $A$ (a) and $\varphi$ (b) images and corresponding line profiles along the $x$ direction. We obtained the line profiles by taking the average of the profiles over the dashed rectangles of the sample. Here, $x = 0$ in the profiles is determined from the peak position of $A$ in (a). The solid orange line in (b) represents the result of the linear fitting. The $x$-directional line profiles in R1 and R2 were fitted by linear functions. (c) $f$ dependence of $D$ measured by heat-current-modulated LIT. The dashed line in (c) represents the value of $D$ obtained by the conventional methods. (d), (e) $A_J$ (d), $\varphi_J$ (e) images and corresponding line profiles along the $x$ direction. (f) $f$ dependence of $\rho c_p$ measured by charge-current-modulated LIT. To estimate $\rho c_p$, we used the averaged $A_J$ values over the dashed rectangle area in (d). The dashed line in (f) represents the values of $\rho c_p$ obtained by the conventional methods.

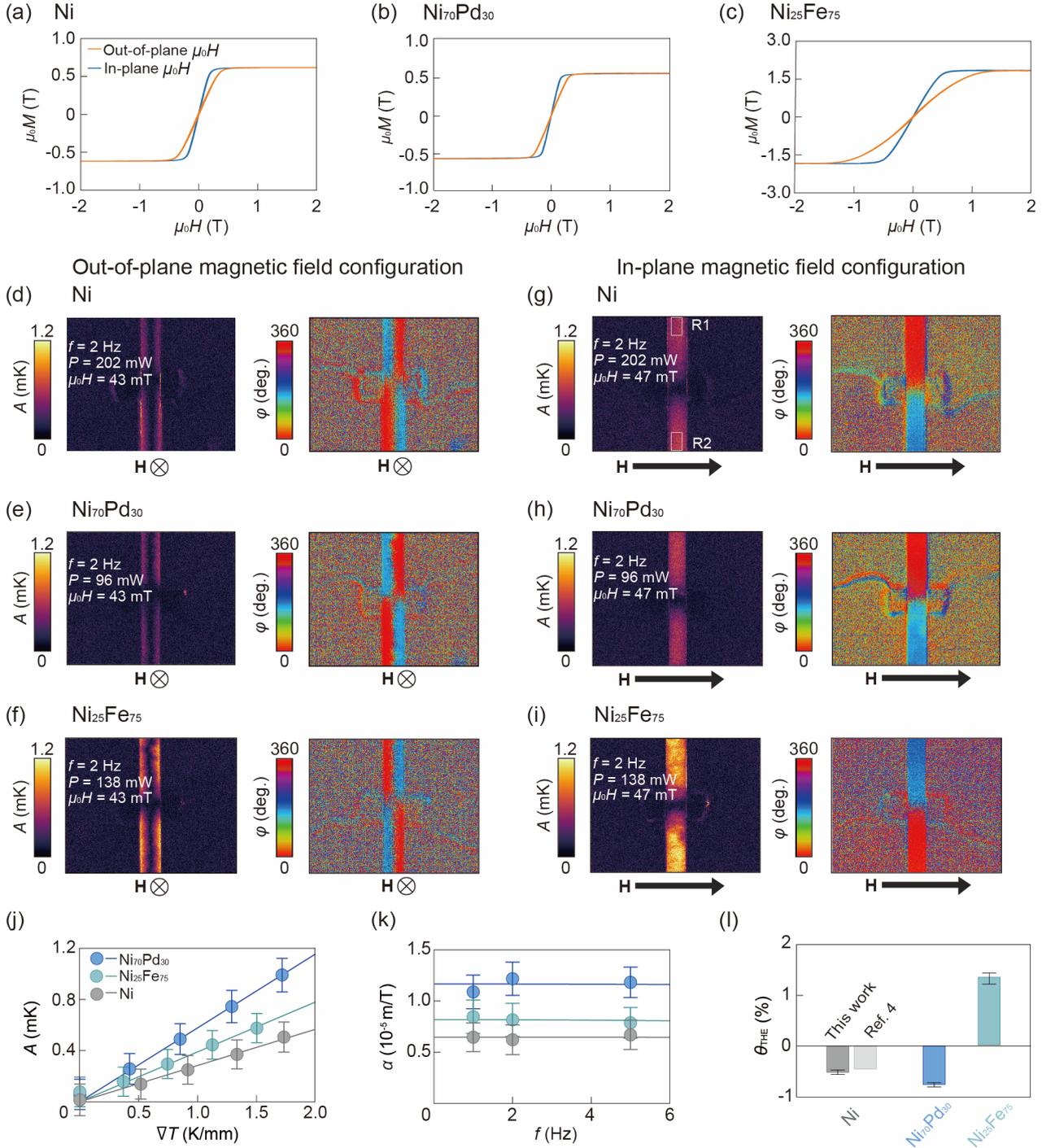

**FIG. 6** (a)-(c) $M$ curves of the Ni, Ni$_{70}$Pt$_{30}$, and Ni$_{25}$Fe$_{75}$ slabs in the out-of-plane and in-plane magnetic field configurations. The magnetic field was applied along the 0.5 mm (1.0 mm) direction of the slabs in the out-of-plane magnetic field (in-plane magnetic field) configurations. (d)-(i) $A$ and $\varphi$ images of the Ni, Ni$_{70}$Pt$_{30}$, and Ni$_{25}$Fe$_{75}$ slabs. The LIT images were obtained at $\mu_0 H_{1f}$ = 43 (47) mT and $f$ = 2.0 Hz in the out-of-plane magnetic field (in-plane magnetic field) configurations, and the heater power was set at $P$ = 202 mW for Ni, 96 mW for Ni$_{70}$Pt$_{30}$, and 138 mW for Ni$_{25}$Fe$_{75}$. The regions R1 and R2 are defined in (g). (j) $\nabla T$ dependence of $A$ in R1 in the in-plane magnetic field configuration at $f$ = 2.0 Hz and $\mu_0 H_{1f}$ = 47 mT. (k) $f$ dependence of $\alpha$. The solid lines show the calculated $f$ dependence of $\alpha$ based on the one-dimensional heat diffusion. (l) $\theta_{THE}$ of Ni, Ni$_{70}$Pt$_{30}$, and Ni$_{25}$Fe$_{75}$ estimated by the LIT-based methods and the literature value of Ni [4].

18